\documentclass[preprint,12pt,3p]{elsarticle}
\usepackage{psfrag,latexsym,subcaption}
\usepackage{graphics} 
\def\1{{1\kern-.3468em{ 1}}}
\usepackage[active]{srcltx}
\usepackage[color,final]{showkeys}

\usepackage{natbib}\usepackage[normalem]{ulem}
\usepackage[english, brazil]{babel}
\usepackage[utf8]{inputenc}
\usepackage{amsmath}
\usepackage{float}
\usepackage{amssymb}
\usepackage{epsfig}
\usepackage{tikz}
\usetikzlibrary{arrows,positioning}
\usepackage{wrapfig}

\usepackage{amssymb}

\usepackage[nodots]{numcompress}

\journal{}

\begin{document}
\selectlanguage{english}

\begin{frontmatter}


\cortext[cor1]{Corresponding author: +44 023 8059 7677}

\title{Reinfection and low cross-immunity as drivers of epidemic resurgence under high seroprevalence: a model-based approach with application to Amazonas, Brazil}

\author[soton]{Edilson F. Arruda \corref{cor1}}
\address[soton]{Department of Decision Analytics and Risk, Southampton Business School, University of Southampton,  Building 2, 12 University Rd, Highfield, Southampton SO17 1BJ, UK }
\ead{e.f.arruda@southampton.ac.uk}
\author[Cefet]{Dayse H. Pastore}
\ead{dayse.pastore@cefet-rj.br}
\author[FRURJ]{Claudia M. Dias} 
\ead{mazzaclaudia@gmail.com}
\author[UFSC]{Fabr\'icio O. Ourique}
\ead{fabricio.ourique@ufsc.br}



\address[Cefet]{Centro Federal de Educação Tecnológica Celso Suckow da Fonseca, Av. Maracanã, 229, Rio de Janeiro, RJ 20271-110, Brasil}
\address[FRURJ]{Graduate Program in Mathematical and Computational Modeling, Multidisciplinary Institute, Federal Rural University of Rio de Janeiro, Av. Gov. Roberto Silveira, s/n - Moqueta, Nova Iguaçu  RJ 26020-740, Brasil}
\address[UFSC]{Department of Computation, Federal University of Santa Catarina, Campus Ararangu\'a, R. Gov. Jorge Lacerda, 3201, Araranguá - SC, 88906-072, Brasil}

\begin{abstract}
This paper introduces a new multi-strain epidemic model with reinfection and cross-immunity to provide insights into the resurgence of the COVID-19 epidemic in an area with reportedly high seroprevalence due to a largely unmitigated outbreak: the state of Amazonas, Brazil. Although high seroprevalence could have been expected to trigger herd immunity and prevent further waves in the state, we have observed persistent levels of infection after the first wave and eventually the emergence of a second viral strain just before an augmented second wave. Our experiments suggest that the persistent levels of infection after the first wave may be due to reinfection, whereas the higher peak at the second wave can be explained by the emergence of the second variant and a low level of cross-immunity between the original and the second variant. Finally, the proposed model provides insights into the effect of reinfection and cross-immunity on the long-term spread of an unmitigated epidemic.
\end{abstract}


\end{frontmatter}


\section{Introduction \label{sec:intro}}

Caused by the new coronavirus (Sars-CoV-2), the COVID-19 pandemic is still ongoing. Reported cases of reinfection evince that acquired immunity wanes over time \cite{Tillett_2020}. While cross-immunity between Sars-CoV-2 and other types of coronavirus can stimulate immune response \cite{Shrockeabd4250}, new variants may be able to evade immunity acquired in previous infections \cite{Greaney2021}. This highlights the importance of understanding and modelling cross-immunity between distinct pairs of viral strains. 

After initially containing severe coronavirus cases, many countries had a resurgence of COVID-19 consistent with a large proportion of the population remaining susceptible to the virus after the first epidemic wave \cite{Sabino2021}. While this is compatible with mitigated outbreaks, it does not conform with the largely unmitigated epidemic in the state of Amazonas (Brazil), which led to a high prevalence after the first wave \cite{Buss2021}. Whilst the high prevalence have led to speculations regarding herd immunity \cite{Taylorn2021}, the epidemic in Amazonas maintained persistent levels of infection after the first wave \cite{Sabino2021}.  Significantly, the discovery of a second viral strain in the state \cite{Faria2021} (variant Gamma) coincided with the second wave of the epidemic, suggesting a correlation between the resurgence and variant Gamma.

Given the inconsistency and possible underestimation of official epidemiological reports in Brazil and elsewhere \cite{Silva2020,Dyern2021}, data-driven approaches do not suffice this phenomenon, as they might reproduce the underlying biases. Indeed, modelling approaches, an understanding of the local reality and a critical analysis of different medical studies are essential to understand the multiple waves of COVID-19 across the globe. This paper proposes a new model of epidemic spread and uses it to search for possible drivers of the COVID-19 resurgence in the state of Amazonas. The model includes reinfection due to waning immunity, multiple virus strains and cross-immunity between pairs of strains.

We find that the resurgence in Amazonas is compatible with a combination of low cross-immunity between the original COVID-19 strain and the emerging Gamma variant, as well as waning immunity. These two possibilities have been recently proposed as possible explanations for the resurgence despite the high prevalence of the disease in the state \cite{Sabino2021}. Our results also suggest that low cross immunity and waning immunity may suffice to explain a higher peak in the second wave, without the need for the second strain to be more transmissible.  We also find that a more transmissible second strain does not suffice to explain a higher second wave when infection from the first strain automatically grants immunity to the second strain. This underscore the important effect of imperfect cross immunity on the magnitude of the second wave.

The state of Amazonas is an interesting case study because the epidemic has been largely unmitigated \cite{Buss2021} and can therefore provide a valid ground for an analysis of the underlying properties of COVID-19's spread under multiple strains. A similar analysis in other realities would require the model to discount the efficacy of social isolation and other mitigating measures, which is a challenging task in itself.

\section{Methods \label{sec:methods}}

To investigate the spread of multiple strains, we propose the epidemic model in \eqref{Eq1-1}-\eqref{Eq4}. Let $V = \{1, \, \ldots, \, n\}$ be the set of virus strains circulating in the population, and $j \in V$ denote a particular strain. For each $j \in V$ and time $t \ge 0$, let $S_j(t), \, E_j(t), \, I_j(t)$ and $R_j(t)$, respectively denote the number of susceptible, exposed, infected and removed (recovered and immune) individuals in the population at time $t$. In addition, $P(t)$ denotes the total population at time $t \ge 0$. $S_j(t)$ includes individuals not immune to strain $j \in V$ at time $t \ge 0$; $E_j(t)$ counts individuals recently contaminated by strain $j$ but still in the latency period - they have not yet manifested the disease; $I_j(t)$ comprises individuals currently carrying strain $j$ in the infectious stage; $R_j(t)$ denotes individuals that are recovered from and immune to strain $j$ at time $t$. The system's dynamics is as follows:

\begin{align}
\dot P(t) &= -\sum_{j=1}^n \mu_j I_j \label{Eq1-1}\\
S_j(t) &= P(t) - E_j(t) - I_j(t) - R_j(t) \label{Eq4-1}\\
\dot{E_{j}}(t) &= \beta_{j} S_{j}(t) I_{j}(t) - \sigma_{j}E_{j}(t), \label{Eq2}\\
\dot{I_{j}}(t) &= \sigma_{j} E_{j} (t) -(\mu_{j}+\gamma_j)I_{j}(t), \label{Eq3} \\
\dot{R_{j}}(t) &= \gamma_jI_{j}(t) + \sum_{i=1, i \ne j}^n c_{ij} \gamma_i I_{i}(t)- \delta_j R_j(t), \label{Eq4}
\end{align}
where $S_j(0) \le P(0), \, \forall \; j \in V$, and $C=[c_{ij}]$ is a symmetric cross-immunity matrix with $0 \le c_{ij} \le 1, \, \forall i,j \in V$. Matrix $C$ is symmetric as the model considers that immunity from strain $i$ extends to strain $j$ in the same way that immunity from strain $j$ extends to strain $i$. 

Table \ref{tab:parameters} conveys the model parameters. In the model, new expositions to strain $j$ are the first term of the right hand side of  \eqref{Eq2}, whereas new infections are the second (resp. first) term in the right hand side of Eq. \eqref{Eq2} (resp. Eq. \eqref{Eq3}). Infectious individuals recover at rate  $\gamma_j > 0$, or die at rate $\mu_j \ge 0$. The ones that recover enter the \emph{removed} compartment in Eq. \eqref{Eq4}; they leave after they immunity wanes at rate $\delta_j$, returning to the susceptible population. This is a novelty compared to classical SEIR epidemic models. Another novelty is the second term in the right-hand side of \eqref{Eq4}. It represents individuals that become immune to strain $j$ immediately after having recovered from strain $i$. This happens due to a cross-immunity factor $0 \le c_{ij} \le 1$: $c_{ji}=0$ indicates no cross immunity, whereas when $c_{ij}=1$ all individuals that recover from strain $i$ also become immune to strain $j$. Hence, we can say that a proportion $c_{ij}$ of individuals that recover from strain $i$ also become immune to strain $j$.

\begin{table}[!ht]
\caption{Parameters for multi-strain dynamics. \label{tab:parameters}}
\centering
\scriptsize{
\begin{tabular}{|c|l|c|} \hline \hline
\textbf{Parameter} & \textbf{Description} & \textbf{Unit} \\ \hline
$\beta_j$ & Transmission rate of strain $j$ & transmissions/encounter \\ \hline
$\sigma_j$ & Inverse of the latency period of strain $j$ & $\text{days}^{-1}$ \\ \hline
$\gamma_j$ & Recovery rate for strain $j$ & $\text{days}^{-1}$ \\ \hline
$\delta_j$ & Rate of immunity loss for strain $j$ & $\text{days}^{-1}$ \\ \hline
$\mu_j$ & Death rate due to strain $j$ & $\text{days}^{-1}$ \\ \hline
$c_{ij}$ & Cross immunity rate between strains $i$ and $j$ & - \\ \hline
\end{tabular}
}
\end{table}

Using a reproduction number $R_0=3$ \cite{Fontanet2020}, we applied the proposed model to the COVID-19 outbreaks in Amazonas and used a serological study \cite{Buss2021} to validate the results. We performed several experiments to provide an insight of the joint effect of cross-immunity and waning immunity (reinfection) when a second viral strain appears after the first strain has stabilised, as the data suggests is the case in the state of Amazonas \cite{Sabino2021}. The parameters and initial conditions used in the experiments appear in Table \ref{tab:parex1}. Parameters $\sigma_j$ and $\beta_j$ ($j=1, 2$) are consistent with studies on COVID-19's latency and infectious periods \cite{Backer2020,Verity2020} and, together with $\beta_j$ in Table \ref{tab:parex1}, produce $R_0=3$ \cite{Fontanet2020}. Parameters $\delta_j$ ($j=1,2)$ were manually fitted to produce a behaviour resembling Amazon's outbreak, whereas $\mu_j$ are in line with the estimates in \cite{Tarrataca2021}.

\begin{table}[htb]
\caption{Parameters for the Experiments \label{tab:parex1}}
\scriptsize{
\centering
\begin{tabular}{|c|c|} \hline \hline
\textbf{Parameter} & \textbf{Value} \\ \hline
$\beta_1=\beta_2$ &  $2.41 \cdot 10^{-9}$ \\ \hline
$\sigma_1=\sigma_2$ & $\frac{1}{7} \, \text{days}^{-1}$ \\ \hline
$\gamma_1=\gamma_2$ & $\frac{1}{21} \, \text{days}^{-1}$ \\ \hline
$\delta_1=\delta_2$ & $\frac{1}{150} \, \text{days}^{-1}$ \\ \hline
$\mu_1=\mu_2$ & $1.152 \cdot 10^{-5} \,\text{days}^{-1}$ \\ \hline
\end{tabular}
\hspace{1cm}
\begin{tabular}{|c|c|} \hline \hline
\multicolumn{2}{|c|}{\textbf{Initial Conditions}} \\ \hline
Strain 1 & Strain 2 \\ \hline
$S_1(0) = 4,144,342$ & $S_2(t) = 4,144,597, \, t < 180$ \\ \hline
$E_1(0) = 252$ & $E_2(t) = 0, \forall t \le 180$, \\ \hline
$I_1(0) = 2$ & $I_2(180) = 1, \, I_2(t) = 0, \forall t < 180$ \\ \hline
$R_1(0) = 1$ &  $\, R_2(t) = 0, \forall t \le 180$ \\\hline
\end{tabular}
}
\end{table}

%
%

\section{Results \label{sec:results}}

\begin{figure}[ht]
\begin{subfigure}{0.5\textwidth}
  \centering
  \includegraphics[width=0.97\linewidth]{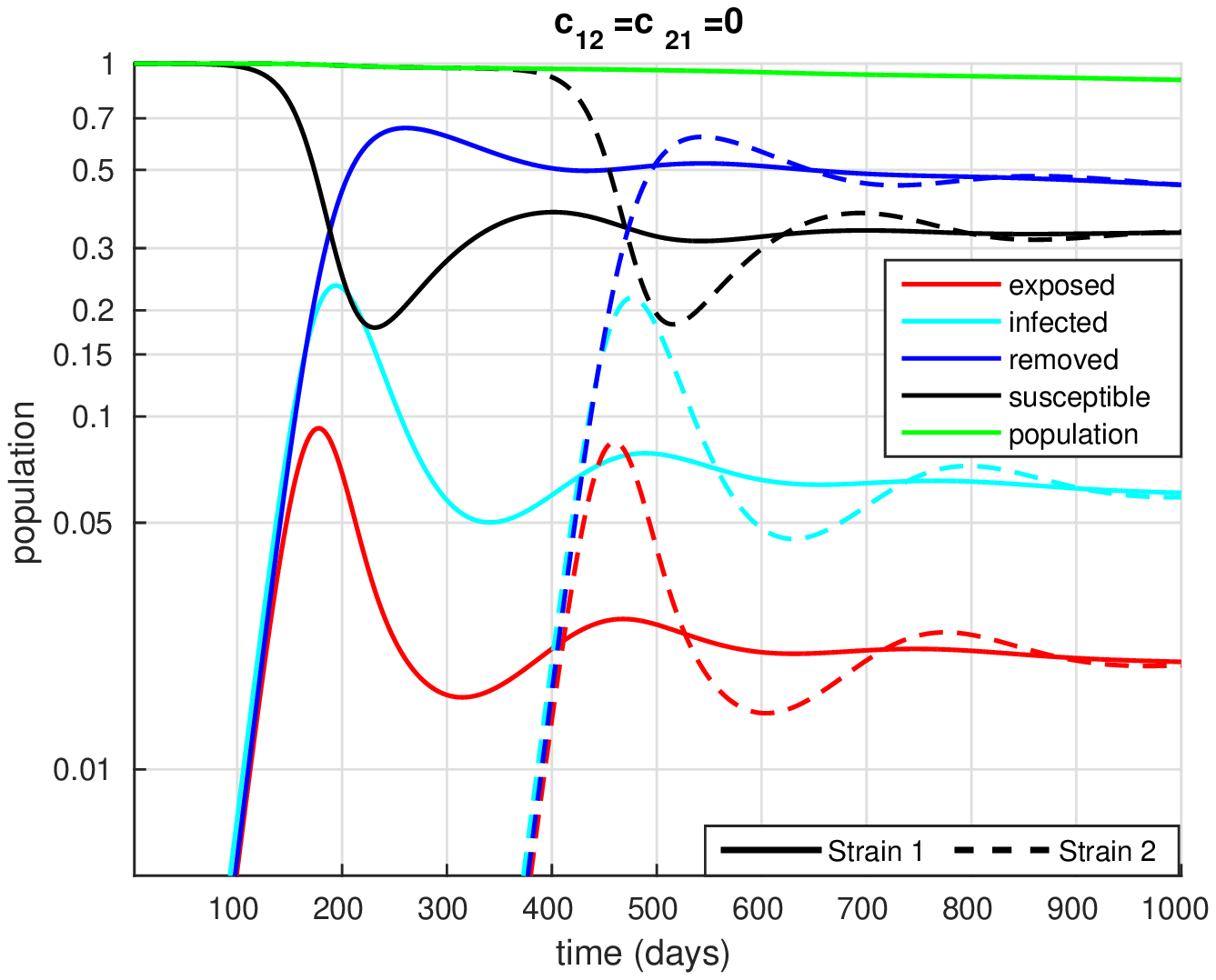}  
  \caption{Two strain dynamics for $c_{12}=0$ \label{fig:cross0}}
\end{subfigure}
\begin{subfigure}{0.5\textwidth}
  \centering
  \includegraphics[width=0.97\linewidth]{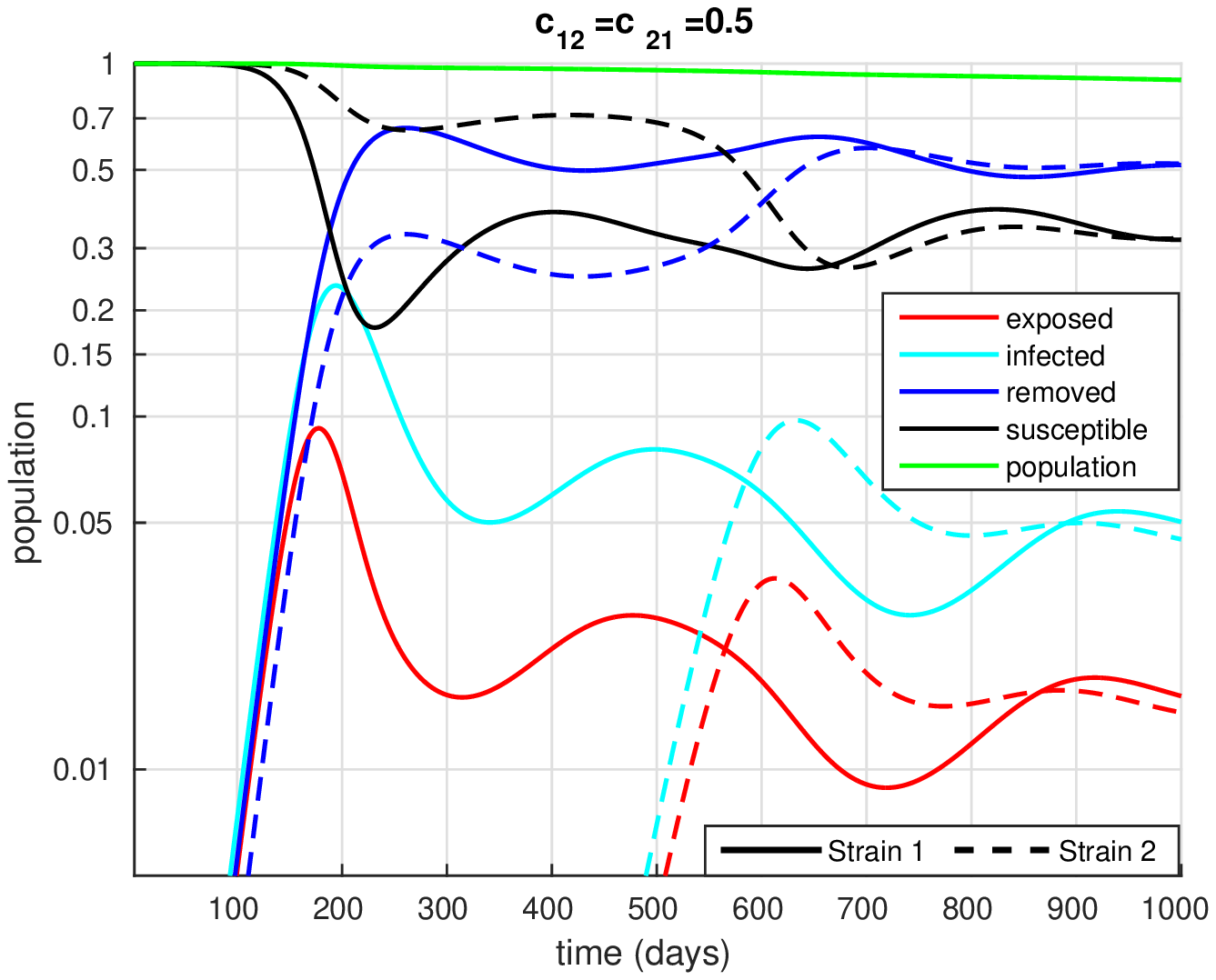}  
  \caption{Two strain dynamics for $c_{12}=0.5$ \label{fig:cross051}}
\end{subfigure}
\caption{Dynamic behaviour for two strains in Amazonas with zero and 50\% cross-immunity \label{fig:extreme}}
\end{figure}

Figure \ref{fig:extreme} depicts the results of two experiments: considering no-cross immunity (Fig. \ref{fig:cross0}), and with 50\% cross-immunity (Fig. \ref{fig:cross051}). We found that the epidemic in Amazonas is consistent with a two-strain outbreak with reinfection and low cross-immunity - as in Fig. \ref{fig:cross0}. Observe that the removed population stabilises around 50\% after the first wave and before the second wave, in line with the prevalence of 52.5\% observed in June 2020 in a seroprevalence study among blood donors in the state \cite{Buss2021}. Whilst the stabilisation after the first outbreak can be explained by reinfection from the first strain, the second peak is due to the emergency of the second strain. As we assumed that the second strain is as transmissible as the first, the second strain is simply a delayed version of the first outbreak. However, at the peak of the second strain we will observe more infections, as these also include cases of the first strain. The number of deaths, observed as a reduction in the population, increases as expected after the emergence of the second strain, since now deaths occur due to both variants.

Figure \ref{fig:cross051} depicts the dynamics when there is a 50\% chance that an infection from strain 1 will trigger immunity for strain 2 and vice-versa, i.e. $c_{12}=c_{21}=0.5$. Observe that the peak of infections is about 10\% for strain 2, which is a significant reduction from the peak of just over 20\% observed with $c_{12}=0$ (Fig. \ref{fig:cross0}). Another important feature is that cross-immunity causes a delay in the second wave. This is to be expected, as cross-immunity produces a reduction in the susceptible population for strain 2 when it appears, thus slowing the spread. Hence, as cross-immunity increases, we can expect reduced and delayed peaks in the second wave.


\begin{figure}[ht]
\begin{subfigure}{0.5\textwidth}
  \centering
  \includegraphics[width=0.97\linewidth]{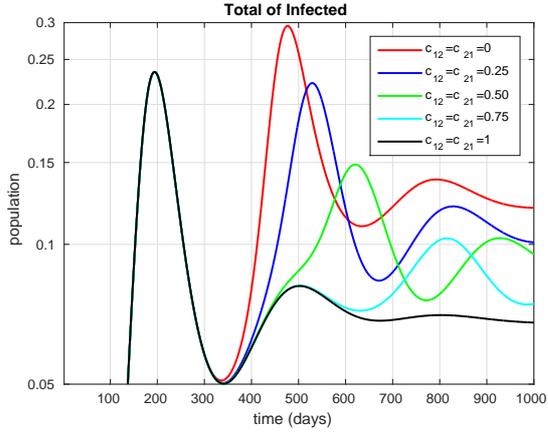}  
  \caption{Infected population for two strains varying $c_{12}$\label{fig:infected}}
\end{subfigure}
\begin{subfigure}{0.5\textwidth}
  \centering
  \includegraphics[width=0.97\linewidth]{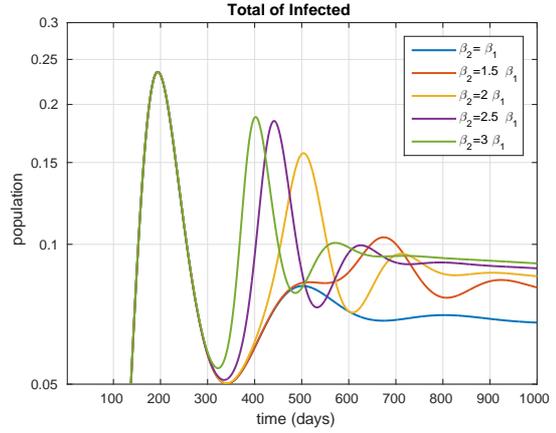}  
  \caption{Infected population for two strains varying $\beta_2$ \label{fig:infected-beta}}
\end{subfigure}
\caption{Two strain dynamics for varying $c_{12}$ and varying $\beta_2$ with $c_{12}=1$}
\label{fig:2strain}
\end{figure}


Fig. \ref{fig:infected} conveys the number of infected individuals across time for varying cross-immunity factors. Under no cross-immunity, the second peak of infections is higher than the first, even though the second strain is as transmissible as the first. This happens because there will be cases from both strains even as the second strain is peaking. There will be more infections due to the second strain at the second peak simply because the peak of the second strain is delayed and happens after the first strain has stabilised. Observe that, as cross-immunity grows, the second peak will be delayed and reduced. It reduces from 30\% when $c_{12}=0$ to approximately $8\%$ when $c_{12}=1$. When $c_{12}=1$, however, the second strain does not cause a second wave, as the removed population for strain 2 has already attained high levels compatible with herd immunity - due to the high levels of infection during the unmitigated first wave. This  illustrates the powerful effect of cross-immunity in the dynamics, which is particularly significant in unmitigated epidemics. Indeed, the results indicate that the role of reinfection in the evolution of the disease may not be as significant as the presence of new strains under the effect of cross-immunity.

To isolate the effect of higher transmissibility, Figure \ref{fig:infected-beta} depicts the effect of higher transmissibility levels for the second strain under full cross-immunity. As we can see, the second peak increases and is anticipated as $\beta_2$ grows. When both strains are equally transmissible, the second strain does not cause a second wave. In contrast, the second peak reaches only about 50\% of the magnitude of the first (12\%) when the second strain is 50\% more transmissible than the first. Even when the second strain is three times more infectious ($\beta_2=3 \beta_1$), the second peak is still significantly lower than the first. This indicates that higher transmissibility is not an overpowering driver under full cross-immunity, which further highlights the effect of imperfect cross-immunity in the magnitude of the second peak.

\section{Discussion \label{sec:discussion}}

The particular scenario of Amazonas (Brazil) served as motivation for the mathematical model presented in this paper, which contemplates the emergence of new viral strains as well as symmetric cross-immunity among viral strains. A sorological study suggested that the state of Amazonas achieved the so-called herd immunity for the first strain in June 2020 \cite{Sabino2021,Buss2021}. However, the number of cases reported did not subside as might be expected. Instead, it maintained stable until a second strain was detected in the state which caused a second wave, more intense than the first. The model-based study suggests that the persistent levels of infection are due to reinfection, whereas the second peak is due to the emergence of a new strain. The results suggest that considering the reinfection from the outset is important for decision making, as it will allow policy makers to estimate the long-term behaviour under the realistic assumption that immunity will probably not last forever.

The results also show that pursuing herd immunity may be shortsighted, as high levels of infection will increase the probability of appearance of new strains capable of evading the immunity acquired for the original strain. Such strains can generate larger outbreaks, as their peak will add to the burden of the original strain when it has not been completely wiped out. The model shows that the second strain need not be more transmissible to generate a larger second wave. In fact, the magnitude of the wave will depend heavily on the cross immunity between the strains. As we cannot control the cross-immunity levels among strains nor ensure that immunity lasts forever, 
long-term policy making should seek to extinguish new epidemics as soon as possible to avoid persistent levels of infection that may lead to dangerous variants and multiple outbreaks. Furthermore, decision support models need to be realistic to ensure that a policies are not hindered by a myopic evaluation of the system.


For the current COVID-19 pandemic, it is possible that other dangerous variants not yet identified could impact disease progression across the globe. The results serve as a warning to the fact that it is necessary to study and monitor new variants, since the emergence of a strain with low cross-immunity with its preceding counterparts could pose a risk for the control of the pandemic. Further, it could compromise the effect of the current mass vaccination progress worldwide. The results suggest that a swift vaccination policy is vital to contain the COVID-19 epidemic and prevent the emergence of additional dangerous variants. In addition, a thorough study of variants and their cross-immunity levels is also a fundamental step for a comprehensive evaluation of the effects of the vaccination and to inform policy-making, leading to a potential return to lock-down measures to curtail more aggressive variants.

\scriptsize{

}

\end{document}